\documentclass{PoS}

\usepackage{graphicx} % Allows including images
\usepackage{booktabs} % Allows the use of \toprule, \midrule and \bottomrule in tables
\usepackage{epsfig,wrapfig,amssymb,color,amsmath,bm,breqn}
\usepackage{amssymb}
\usepackage{amsmath}
\usepackage{bm}

\usepackage{ulem}
\usepackage[dvipsnames]{xcolor}

\usepackage{cite}

\title{Applications of the WW-type approximation to SIDIS}

\ShortTitle{Applications of the WW-type approximation to SIDIS}

\author{S. Bastami,$^a$ H. Avakian,$^{b}$ A. V. Efremov,$^{c}$ A. Kotzinian,$^{de}$ B. U. Musch,$^{f}$ \newline B. Parsamyan,$^{g}$ A. Prokudin,$^{h}$ M. Schlegel,$^{i}$ G. Schnell,$^{j}$ P. Schweitzer$^{a}$\newline and \speaker{K. Tezgin}$^{a}$\\
             \llap{$^a$}Department of Physics, University of Connecticut,
        Storrs, CT 06269, U.S.A.\\
             \llap{$^b$}Thomas Jefferson National Accelerator Facility,
        Newport News, VA 23606, U.S.A.\\
             \llap{$^c$}Joint Institute for Nuclear Research, Dubna,
             141980 Russia \\
             \llap{$^d$}Yerevan Physics Institute,  Alikhanyan Brothers St.,
        375036 Yerevan, Armenia\\
             \llap{$^e$}INFN, Sezione di Torino,
        10125 Torino, Italy\\
             \llap{$^f$}Institut f\"ur Theoretische Physik, Universit\"at
        Regensburg, 93040 Regensburg, Germany\\
             \llap{$^g$}CERN, 1211 Geneva 23, Switzerland\\
             \llap{$^h$}Division of Science, Penn State Berks, Reading,
        PA 19610, USA\\
             \llap{$^i$}Department of Physics, New Mexico State University,
        Las Cruces, NM 88003-001, USA\\
             \llap{$^j$}Department of Theoretical Physics, University of the Basque
        Country UPV/EHU, 48080 Bilbao, Spain, and
        IKERBASQUE, Basque Foundation for Science, 48013 Bilbao, Spain\\
             E-mail:  \email{saman.bastami@uconn.edu}, \email{avakian@jlab.org},
                      \email{efremov@theor.jinr.ru}, \email{aram.kotzinian@cern.ch},
                      \email{bmusch@b-mu.de}, \email{bakur@cern.ch},
                      \email{prokudin@jlab.org}, \email{schlegel@nmsu.edu},
                      \email{gunar.schnell@desy.de}, \email{peter.schweitzer@phys.uconn.edu},
                      \email{kemal.tezgin@uconn.edu}}

\abstract{We explore the complete cross-section for the production of unpolarized hadrons in semi-inclusive deep-inelastic scattering up to power-suppressed $\mathcal{O}(1/Q^2)$ terms in the Wandzura-Wilczek-type (WW-type) approximation, which consists in systematically assuming that $\bar{q}gq$-correlators are much smaller than $\bar{q}q$-correlators. Under the applicability of WW-type approximations, certain relations among transverse momentum dependent parton distribution functions (TMDs) and fragmentation functions (FFs) occur which are used to approximate SIDIS cross-section in terms of a smaller subset of TMDs and FFs. We discuss the applicability of the WW-type approximations on the basis of available data.}

\FullConference{XXVII International Workshop on Deep-Inelastic Scattering and Related Subjects - DIS2019\\
		8-12 April, 2019\\
		Torino, Italy}

\begin{document}

\section{Introduction}
%TMDs and FFs contain a wealth of information about 3-dimensional structure of the nucleon.
TMDs and FFs are one of the main ingredients to study the 3-dimensional structure of the nucleon. Although they are well defined matrix elements in QCD, their accessibility through experiments is challenging. One powerful tool to study TMDs and FFs is the semi-inclusive deep-inelastic scattering (SIDIS) process. The SIDIS cross section consists of 18 structure functions (SFs) \cite{Kotzinian:1994dv,Mulders:1995dh,Bacchetta:2006tn} which at leading twist can be expressed by a convolution over transverse momenta of a TMD $f$ and a FF $D$:
\begin{equation}\label{Eq:Convolution}
\mathcal{C}\biggl[\omega\;f\;D\biggr]=x \sum_a e_a^2\int d^2{\mathbf{k}}_\perp d^2{\mathbf{P}}_\perp\delta^{(2)}(z \mathbf{k}_\perp+ \mathbf{P}_\perp-\mathbf{P}_{hT})\;\omega f^a(x,k_\perp^2)\ D^a(z,P_\perp^2),
\end{equation}
with a weight function $\omega$ and at subleading twist, under the validity of factorization, can be expressed by a superposition of such convolutions. However, due to limited knowledge of higher-twist TMDs and FFs at hand, one might want to investigate useful relations among TMDs (FFs). WW-type approximation may be useful for this purpose. The approximation has first been established for the twist-3 PDF $g_T(x)$ \cite{Wandzura:1977qf} and later for $h_L(x)$ \cite{Jaffe:1991ra}. By using the QCD equations of motion, the operators defining those PDFs can be decomposed in terms of a twist-2 $\bar{q}q$-term, a twist-3 $\bar{q}gq$-term and a current-quark mass term. We denote the last two collectively by functions with a tilde. The tilde terms in $g_T(x)$, $h_L(x)$ were predicted in the instanton vacuum model \cite{Balla:1997hf,Dressler:1999hc} to be small compared to $\bar{q}q$-terms. Neglecting the tilde terms is commonly known as WW approximation and is also supported by lattice results \cite{Gockeler:2000ja,Gockeler:2005vw}. The experimental data for $g_T$ show that the approximation works within an accuracy of $40\%$ or better \cite{Accardi:2009au}.

\section{WW-type approximation for TMDs and FFs}
Using QCD equations of motion, one can decompose twist-3 TMDs and FFs into $\bar{q}q$ and tilde terms. WW-type approximation generalizes the idea of WW-approximation to TMDs by assuming $|\langle \bar{q}gq \rangle | \ll | \langle \bar{q}q \rangle |$. Since in TMDs and FFs we deal with unintegrated $\bar{q}gq$-correlations, we distinguish the approximation from the WW approximation and rather refer to them as WW-type approximation. As a result of this approximation, all twist-2 and twist-3 SIDIS SFs can be expressed in terms of 8 leading twist basis functions which include 6 TMDs ($f_1,g_1,h_1,f_{1T}^\perp,h_1^\perp,h_{1T}^\perp$) and 2 FFs ($D_1, H_1^\perp$). Combining WW and WW-type approximations along with a Gaussian Ansatz for transverse momentum dependence of the TMDs and FFs \footnote{The Gaussian widths of TMDs and FFs are denoted by $\langle k_\perp^2 \rangle_{TMD}$ and $\langle P_\perp^2 \rangle_{FF}$, respectively.}, enables us to calculate all SIDIS SFs in terms of the 8 basis functions. Two of such relations are worth mentioning here \cite{Bastami:2018xqd}:
\begin{subequations}\begin{alignat}{3}
   	x \; g_T^a(x) &=&
        \phantom{2} x \;\int_x^1\frac{ d y}{y}\,g_1^a(y) + &x \; \tilde{g}_T^a(x)
        \stackrel{\rm WW}{\approx}
        \phantom{x} x \;\int_x^1\frac{ d y}{y}\,g_1^a(y) \; 
        \stackrel{\rm WW-type}{\approx} \;
        \frac{\langle k_\perp^2 \rangle_{g_T}}{2\;M_N^2} \; g_{1T}^{\perp a}(x),
	\label{Eq:WW-original1} \\
   	x \; h_L^a(x) &=& \; 2x^2\int_x^1\frac{ d y}{y^2}\,h_1^a(y) + &x \; \tilde{h}_L^a(x)
        \stackrel{\rm WW}{\approx} 
        2x^2\int_x^1\frac{ d y}{y^2}\,h_1^a(y)\;
        \stackrel{\rm WW-type}{\approx} \;
        -\frac{ \langle k_\perp^2 \rangle_{h_L}}{M_N^2} \; h_{1L}^{\perp a}(x).
	\label{Eq:WW-original2}
\end{alignat}\end{subequations}
 Exploratory investigations of TMDs on the lattice \cite{Musch:2010ka,Hagler:2007xi,Gockeler:2005cj}, can help us to test the WW-type approximations. In these early works the transverse momentum dependent observables are not precisely those measurable in SIDIS and the Drell-Yan process. Nevertheless, these lattice studies indicate that certain WW-type approximations are satisfied for the lowest Mellin moments as discussed in \cite{Bastami:2018xqd}. Besides, the lattice results also support the Gaussian Ansatz \cite{Musch:2010ka}. Many WW-type relations hold also in quark models \cite{Metz:2008ib}. We use state-of-the-art parametrizations for the basis functions. Our goal is to examine the applicability of WW-type approximation with current available data. Our results will help to deepen the understanding of how sizable the $\bar{q}gq$-correlations are.

\section{Example of leading-twist asymmetry in WW-type approximation}
To test the applicability of the approximations, we studied all the spin and azimuthal asymmetries of SIDIS up to $1/Q^2$ accuracy. We used the following definition for spin asymmetries:
\begin{equation}
    A_{XY}^w(x,z,P_{hT})=\frac{F_{XY}^w(x,z,P_{hT})}{F_{UU,T}(x,z,P_{hT})},
\end{equation}
where X (U/L) and Y (U/L/T) denote the beam and the target polarizations and $w$ is the azimuthal-angle modulation for the corresponding structure function. In some cases, based on the available experimental data, we needed to include kinematic prefactors in the definition of the asymmetry (denoted by subscript $\langle y\rangle$). The $F_{UU,T}$ structure function is obtained in terms of the unpolarized TMD and FF, where we used \cite{Martin:2009iq,deFlorian:2007aj,Anselmino:2005nn} for numerical values. By integrating over $P_{hT}$ we get 
\begin{equation}
    F_{UU,T}(x,z)=x \sum_a e_a^2 \ f_1^a(x) \ D_1^a(z).
\end{equation}

Two of the leading twist structure functions and all eight subleading ones are amenable to WW-type approximations. In this proceeding we present selected results to illustrate the key features of the approach. For all results and more details we refer to \cite{Bastami:2018xqd}.

\begin{figure}[t!]
\centering
\includegraphics[width=0.3\textwidth]{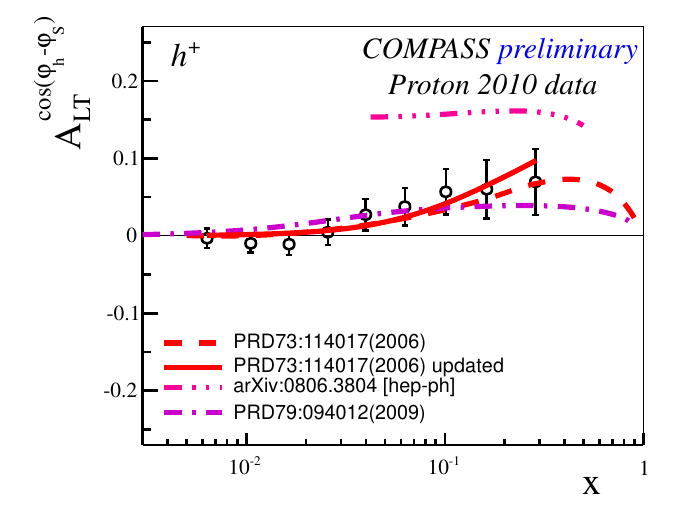} {\tiny (a)}
\includegraphics[width=0.3\textwidth]{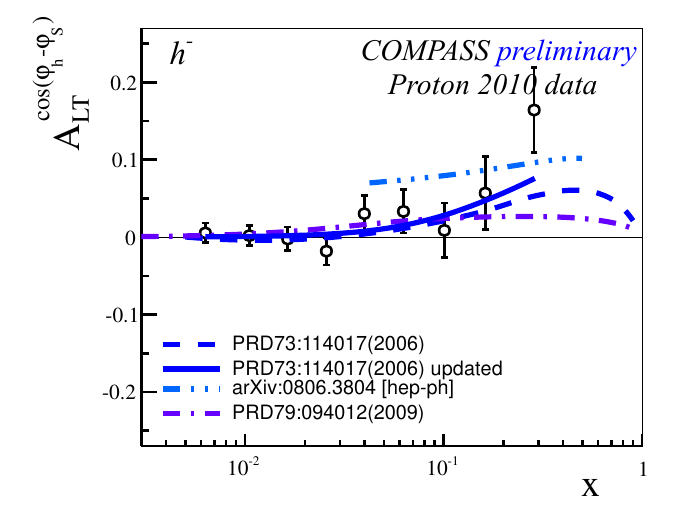} {\tiny (b)}
\includegraphics[width=0.265\textwidth]{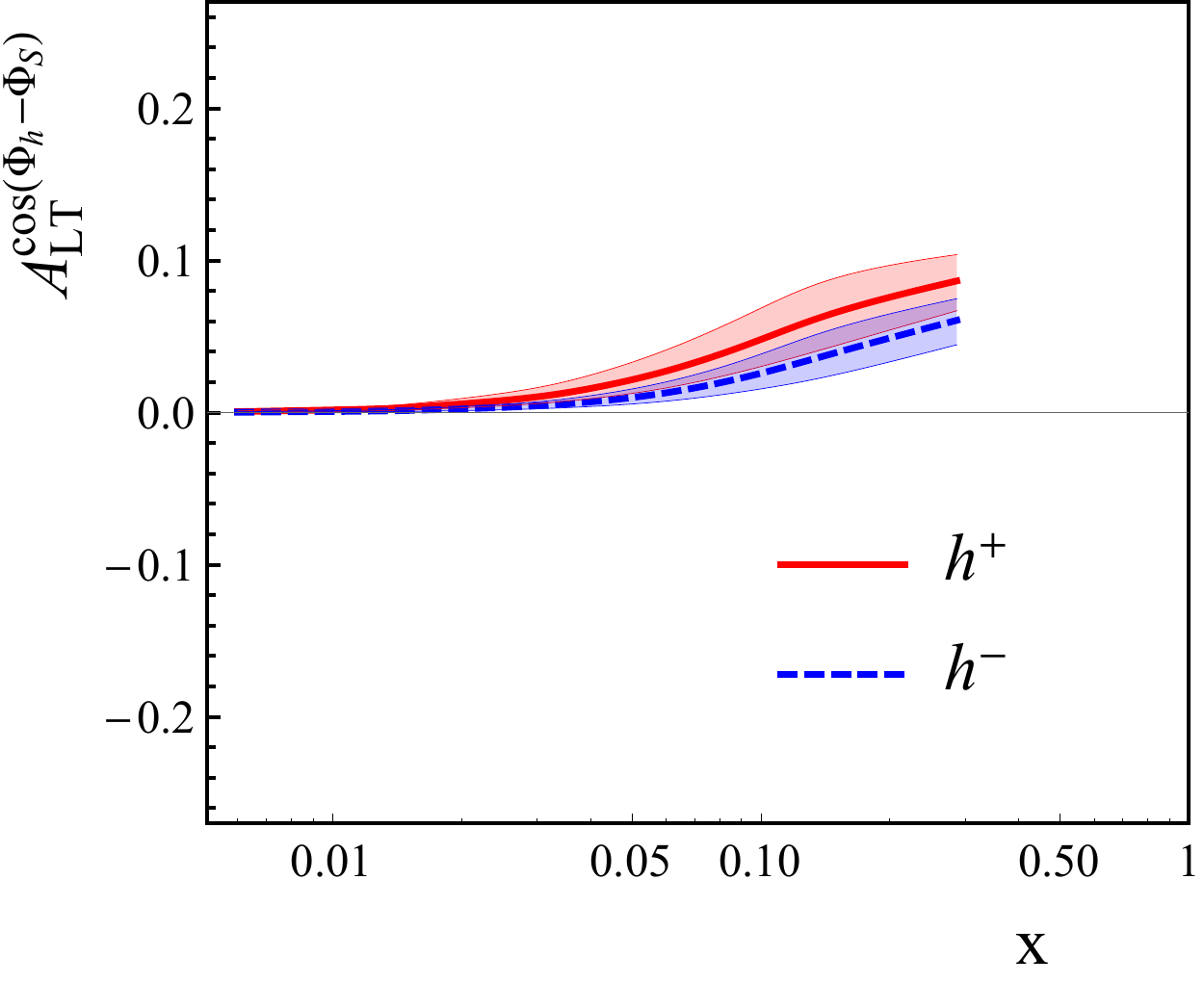} {\tiny (c)}
	\caption{\label{LTcosphihphimphi}
	Preliminary COMPASS data on $A_{LT}^{\cos(\phi_h-\phi_S)}$ \cite{Parsamyan:2013fia} compared with several model predictions \cite{Kotzinian:2006dw, Kotzinian:2008fe, Boffi:2009sh} (a,b), and our calculation for COMPASS kinematics (c).
	}
\end{figure}

As an example of leading twist, we consider the SF $F_{LT}^{\cos(\phi_h-\phi_S)}$ which is a convolution of $g_{1T}^{\perp}$ with the unpolarized FF $D_1$. The WW-type approximation in Eq.~\ref{Eq:WW-original1} is used to calculate $g_{1T}^\perp$ for which we used the parametrization of the collinear $g_1(x)$ from \cite{Gluck:1998xa}. After integrating over the hadron transverse momentum $P_{hT}$, we obtain
\begin{equation}
    F_{LT}^{\cos (\phi_h-\phi_S)}(x,z) = x \sum_a e_a^2 \ g_{1T}^{\perp a}(x) \ D_1^a(z) \ \bigg(\frac{\sqrt{\pi} \ z \  \langle k_\perp^2  \rangle _{g_{1T}^{\perp}}}{2 \ M_N \ \sqrt{\lambda}}\bigg),
\end{equation}
where $\lambda= z^2  \langle k_\perp^2  \rangle _{g_{1T}^\perp}+ \langle P_\perp^2  \rangle _{D_1}$. We assumed $ \langle k_\perp^2  \rangle _{g_{1T}^{\perp}}= \;  \langle k_\perp^2  \rangle _{g_1}$ the value of which is borrowed from Lattice predictions \cite{Hagler:2009mb}. Fig.~\ref{LTcosphihphimphi} shows preliminary COMPASS data for charged hadrons along with our calculations of the asymmetry in the relevant kinematics of the experiment. The results show a good compatibility within the range of uncertainties.

%%%%%%%%%%%%%%%%%%%%%%%%%%%%%%%%%%%%%%%%%%%%%%%%%%%%%%%%%%%%%%%%%%%%%%%%%%%%%

%%%%%%%%%%%%%%%%%%%%%%%%%%%%%%%%%%%%%%%%%%%%%%%%%%%%%%%%%%%%%%%%%%%%%%%%%%%%
\section{Examples of subleading-twist asymmetries in WW-type approximation}
The subleading-twist structure functions are more complex than the leading-twist ones. The WW-type approximation provides us a crucial simplification to describe subleading-twist asymmetries. By applying WW-type approximations in structure functions, one ends up with a smaller number of terms which are expressed in terms of known basis functions. We illustrate our findings with two examples. For more details we refer to \cite{Bastami:2018xqd}.
%\subsection{$A_{LT}^{cos\;\phi_S}$}

\begin{figure}[t!]
\centering
%\begin{tabular}{ccc} 
\includegraphics[height=3.3cm]{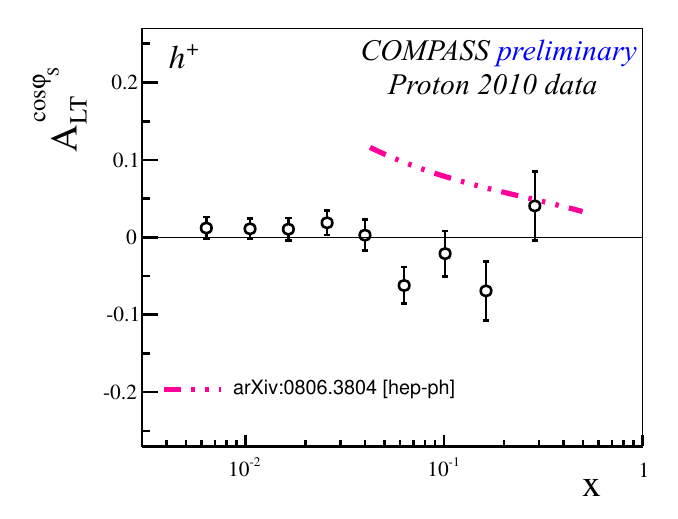} {\tiny (a)}
\includegraphics[height=3.3cm]{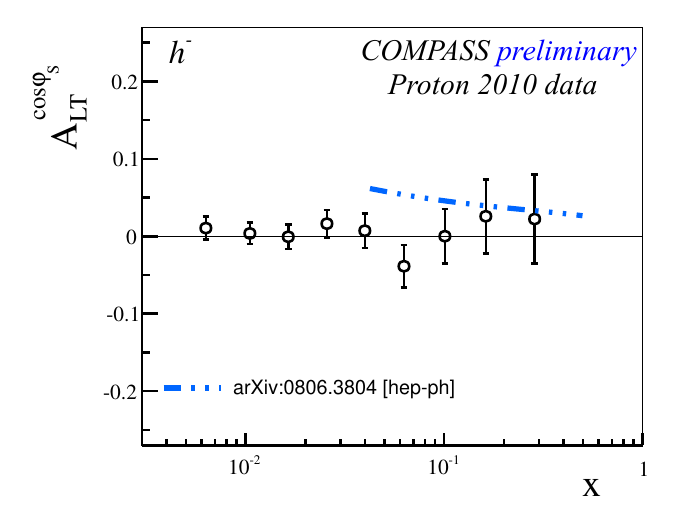} {\tiny (b)}
\includegraphics[height=3.13cm]{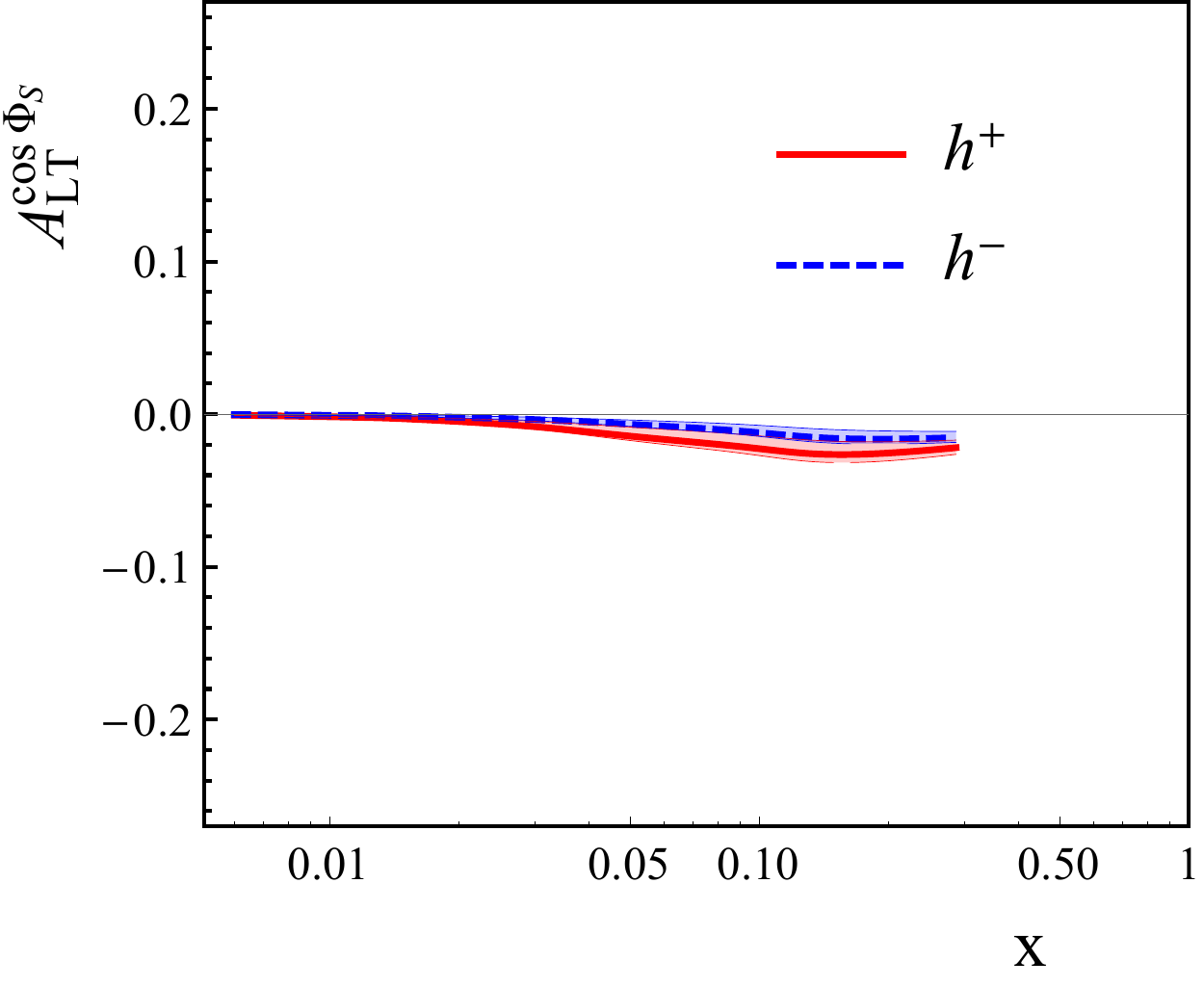} {\tiny (c)}
%{\tiny(a)}& {\tiny(b)}& {\tiny(c)}
%\end{tabular}
    \caption{\label{LTcosphiS}
    Preliminary COMPASS data on $A_{LT}^{\cos(\phi_S)}$ \cite{Parsamyan:2013fia} compared with a model prediction \cite{Kotzinian:2008fe} (a,b), and our calculation for COMPASS kinematics (c).
} 
\end{figure}

The first example is the structure function $F_{LT}^{\cos \phi_S}$ consisting of six terms:
\begin{align}
F_{LT}^{\cos \phi_S}(x,z,P_{hT})=-\frac{2M_N}{Q}&\,{\cal C}\biggl[
        \biggl(x   g_T   D_1
        + \frac{m_h}{M_N}\, h_{1}  \frac{\tilde{E}}{z} \biggr)\nonumber \\
        &-\frac{P_\perp\cdot k_\perp}{2\,z\,M_N\,m_h}
        \biggl( x   e_{T}  H_{1}^{\perp }
        - \frac{m_h}{M_N}\, g_{1T}^\perp \,\frac{\tilde{D}^{\perp }}{z}
        +  x   e_{T}^{\perp }  H_{1}^{\perp }
        + \frac{m_h}{M_N}\,f_{1T}^{\perp }\,\frac{\tilde{G}^{\perp }}{z}\biggr)\biggr].
\end{align}
All but one term vanish after applying the WW-type approximations which leaves us with a convolution of $g_T$ and $D_1$. The $P_{hT}$ integrated structure function then reads
\begin{equation}
F_{LT}^{\cos\phi_S}(x,z)= -\frac{2M_N}{Q}\; x^2 \sum_a e_a^2\, g_T^a(x)\,D_1^a(z).
\end{equation}
The collinear function $g_T^a(x)$ is related to $g_1^a(x)$ (see Eq.~\ref{Eq:WW-original1}) and, therefore, $F_{LT}^{\cos\phi_S}(x,z)$ can be expressed by basis functions. Fig.~\ref{LTcosphiS} shows preliminary COMPASS data for charged hadrons along with our calculations of the asymmetry in the relevant kinematics of the experiment. The predicted asymmetry is small and compatible with the preliminary COMPASS data within uncertainties.
%%%%%%%%%%%%%%%%%%%%%%%%%%%%%%%%%%%%%%%%%%%%%%%%%%%%%%%%%%%%%%%%%%%%%%%%%%

%Subleading-twist asymmetry $A_{LT}^{\cos\phi_S}$ as function of $x$ from scattering of 160 GeV longitudinally polarized muons off a transversely polarized proton target \cite{Parsamyan:2013fia} (a,b), and our calculation for COMPASS (c).
%%%%%%%%%%%%%%%%%%%%%%%%%%%%%%%%%%%%%%%%%%%%%%%%%%%%%%%%%%%%%%%%%%%%%%%%%%%
%\subsection{$A_{UL}^{sin\;\phi_h}$}

\begin{figure}[t]
\centering
\includegraphics[width=0.4\textwidth]{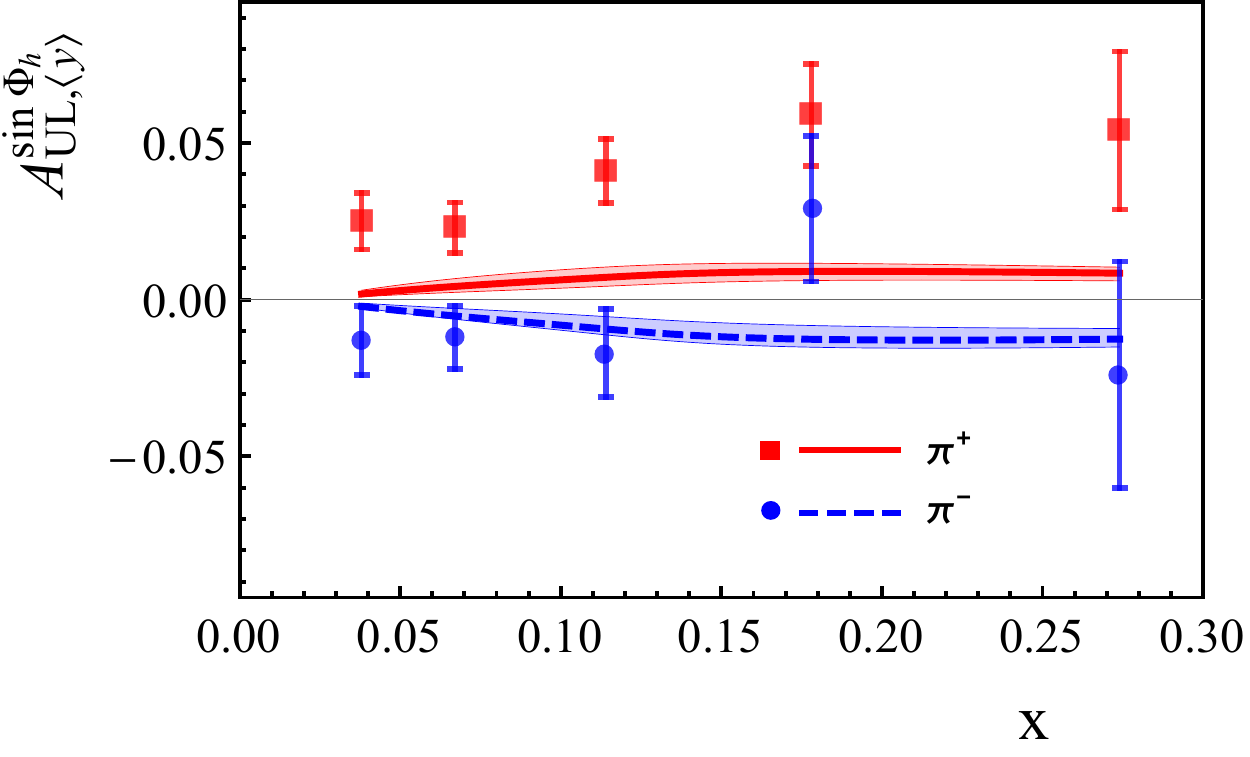}\hspace{1cm}%
\includegraphics[width=0.4\textwidth]{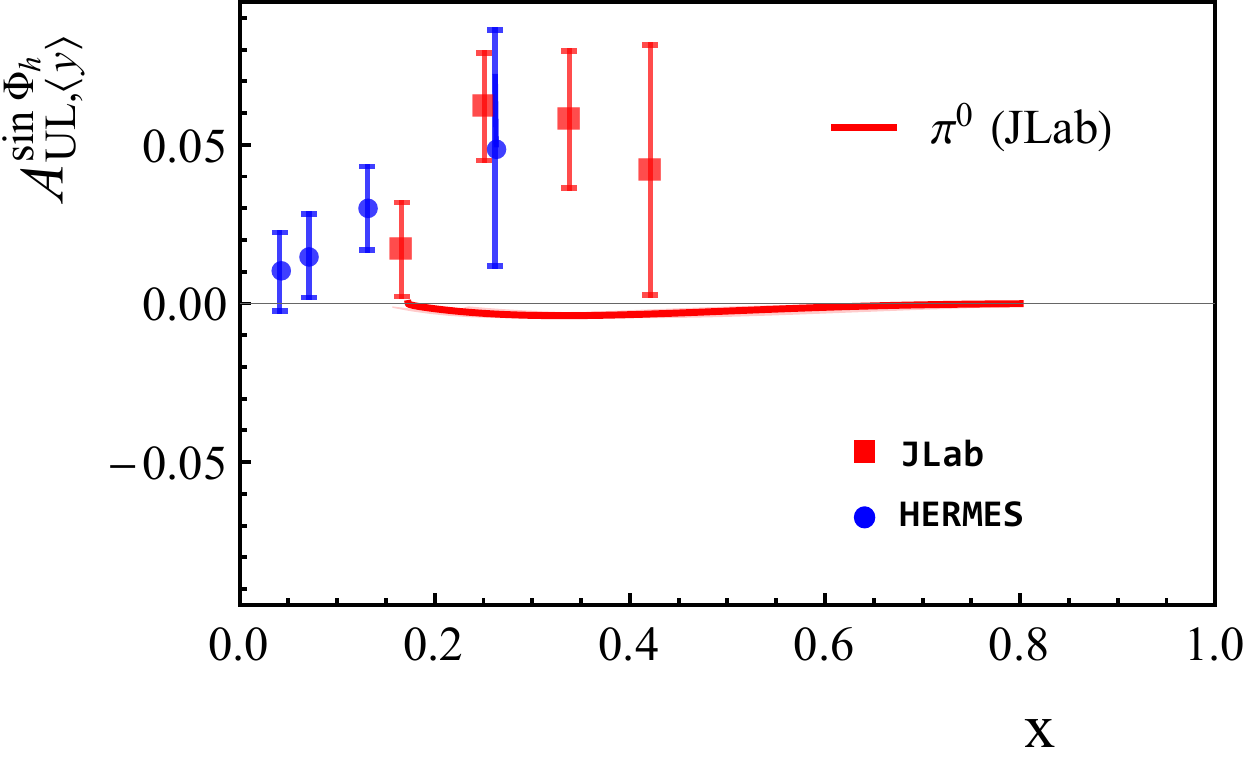} \\
\caption{\label{aulsinphi_jlab}
	$A_{UL}^{\sin\phi_h}$ for proton target vs $x$ from WW-type
	approximation in comparison to data. Left: $\pi^\pm$ from HERMES \cite{Airapetian:2005jc}. Right: $\pi^0$ from HERMES \cite{Airapetian:2001eg}
	and JLab \cite{Jawalkar:2017ube}.
%	Lower panel: preliminary $h^\pm$ COMPASS data \cite{Parsamyan:2018ovx,Parsamyan:2018evv}.
}
\end{figure}

The second example is the structure function $F_{UL}^{sin\;\phi_h}$ consisting of six terms. After implementing the WW-type approximations, the only non-vanishing term in this SF is a convolution of the twist-3 TMD $h_L$ and the Collins FF $H_1^\perp$. Integrating this SF over $P_{hT}$ yields
\begin{equation}
    F_{UL}^{\sin \phi_h}(x,z) = \frac{2 \; M_N}{Q} \ x^2 \sum_a e_a^2 \; h_L^a(x) \ H_1^{\perp a}(z) \ \bigg(\frac{\sqrt{\pi} \  \langle P_\perp^2  \rangle _{H_1^\perp}}{2 \; z \; m_h \; \sqrt{\lambda}}\bigg),
\end{equation}
where $\lambda= z^2  \langle k_\perp^2  \rangle _{h_L}+ \langle P_\perp^2  \rangle _{H_1^\perp}$. We assume that $ \langle k_\perp^2  \rangle _{h_L}\;=\; \langle k_\perp^2  \rangle _{h_1}$ and hence, thanks to WW and WW-type approximations, the collinear part of $h_L$ is related to transversity PDF $h_1$ through Eq.~\ref{Eq:WW-original2}. Parametrizations from \cite{Anselmino:2013vqa} are used for transversity and Collins function. In Fig.~\ref{aulsinphi_jlab} we depict our results for neutral and charged pions. While the WW-type approximation is not incompatible with data for $\pi^-$, it underestimates $\pi^+$ production at HERMES. We face similar situation for $h^{\pm}$ production at COMPASS \cite{Parsamyan:2018ovx,Parsamyan:2018evv}. We also find that the approximation does not explain the large effect for $\pi^0$ production at HERMES and JLab. These results hint at non-negligible contributions from tilde terms.
%%%%%%%%%%%%%%%%%%%%%%%%%%%%%%%%%%%%%%%%%%%%%%%%%%%%%%%%%%%%%%%%%%%%%%%%%%%%
\section{Conclusions}
Our results indicate that WW-type approximation works in certain but not all cases. Further results from \cite{Bastami:2018xqd} are as follows. $A_{UL}^{sin (2\phi_h)}$ shows compatibility with preliminary COMPASS data, however, more precise data is needed to reliably conclude from the comparison. Data for subleading-twist asymmetries $A_{LT}^{cos (2\phi_h-\phi_S)}$, $A_{LL}^{cos \; \phi_h}$ and $A_{UT}^{sin (2\phi_h-\phi_S)}$ are compatible with the approximation. We should note that in most of these cases the asymmetries are measured to be very close to zero. The $A_{UU}^{cos \; \phi_h}$ is overshot by the approximation which calls for more studies of this asymmetry. In WW-type approximation, $A_{UT}^{sin \; \phi_S}$ is predicted to be zero when integrated over $P_{hT}$, while HERMES \cite{Schnell:2010} and COMPASS \cite{Parsamyan:2013fia} data clearly show non-zero effects for $x \gtrsim 0.1$. Another interesting asymmetry is the $A_{LU}^{sin \; \phi_h}$ in which WW-type approximation is not applicable because only $\bar{q}gq$-correlators contribute to this asymmetry. It might be quite interesting to study this particular case to deepen our understanding on $\bar{q}gq$-correlators. In cases where the WW-type approximation does not work, one could study which of the tilde terms neglected in this work, if any, plays a dominant role in the asymmetry.   
\newline

This work was partially supported by NSF PHY-1623454 and PHY-1812423, DOE DE-AC05-06OR23177 and DE-FG02-04ER41309, and the TMD Collaboration framework.

\end{document}